# In Situ X-Ray Radiography and Tomography Observations of the Solidification of Alumina Particles Suspensions

# Part I: Initial Instants

Sylvain Deville[1*], Eric Maire[2], Audrey Lasalle[1], Agnès Bogner[2], Catherine Gauthier[2], Jérôme Leloup[1], C. Guizard[1]

[1] Laboratory of Synthesis and Functionnalisation of Ceramics, UMR3080 CNRS/Saint-Gobain CREE, Cavaillon, France

[2] Université de Lyon, INSA-Lyon, MATEIS CNRS UMR5510, 7 avenue Jean Capelle, F-69621 Villeurbanne, France

Abstract

This paper investigates by in situ high-resolution X-ray radiography and tomography the behaviour of colloidal suspensions of alumina particles during directional solidification by freezing. The combination of these techniques provided both qualitative and quantitative information about the propagation kinetic of the solid/liquid interface, the particles redistribution between the crystals and a particle enriched phase and the 3D organisation of the ice crystals. In this first part of two companion papers, the precursor phenomena leading to directional crystallisation during the first instants of solidification is studied. Mullins-Sekerka instabilities are not necessary to explain the dynamic evolution of the interface pattern. Particle redistribution during these first instants is dependent on the type of crystals growing into the suspension. The insights gained into the mechanisms of solidification of colloidal suspensions may be valuable for the materials processing routes derived for this type of directional solidification (freeze-casting), and of general interest for those interested in the interactions between solidification fronts and inert particles.

*Keywords: directional solidification, X-ray tomography, synchrotron radiation, ceramics, crystal growth*

## 1. Introduction

The solidification of suspensions of particles is of critical importance in various and sometimes extremely different fields, such as the preservation of biological materials [1](cells, food), the processing of composites materials [2] (metals with inclusions, ceramic materials), the freezing of soils [3] or the purification of water from pollutants [4]. Recent developments in the so-called freeze-casting processing routes, in particular for ceramic materials [5] but also for metals [6], have drawn attention on the behaviour of such suspensions undergoing directional solidification.

---

[*] Corresponding author : sylvain.deville@saint-gobain.com

What is understood so far of the freeze-casting process has been analysed mostly from post-mortem observations [7]: cross-section imaging of ceramic scaffolds obtained after complete solidification of the starting suspension, followed by sublimation and densification of the green body by sintering at high-temperature. Two major limitations have hindered further progress in the comprehension of the solidification process. First, the observations are two-dimensional, hence giving only a partial view of the structure or the arrangement of particles. Second, only a static picture of the particles arrangement after solidification is obtained, and this arrangement is also likely to be modified by the sublimation and sintering steps with mass transport occurring at high-temperature. Hence both three-dimensional and dynamic observations of the solidification stage are highly desirable.

In addition, the knowledge regarding the behaviour of the solidification of colloidal suspension has been largely extrapolated from the behaviour of suspension of single and larger particles [8-11], because these can be more easily observed (using conventional optical microscopy techniques for example). The behaviour of suspensions of large particles differs largely from that of colloidal suspensions, at least regarding two aspects: the segregation of particles, which does not occur with colloidal particles where Brownian motion is dominating, and the interaction between the particles and between the particle and the solid/liquid interface. Any conclusion drawn from such experiments with large particles should be extrapolated to colloidal particles with great care.

In situ observation is therefore highly desirable to investigate several unknown aspects of the solidification process, including the morphological evolution of crystal shapes that are growing below the liquid/solid interface during the first steps of solidification and the particles redistribution occurring in front of this interface during crystal growth. Finally, in situ observation allows systems with more simple formulation to be used: when processing porous ceramic samples using the solidification of a colloidal suspension, an organic binder must be added to the formulation, so that the particles arrangement does not collapse during sublimation. However, this binder might affect several aspects of the process, such as the particle's interactions and surface energy, the viscosity of the suspension, the freezing temperature of the media, the diffusion of the species ahead of the interface and so forth. The presence of the binder is not necessary when performing in situ experiments, so that the system is closer to the ideal model systems (particles suspended in water) investigated theoretically [12,13].

X-ray radiography and tomography experiments provide several advantages in this context [14,15]: transparent materials (like those used in [16]) are not required since the imaging is based on absorption contrast; a high spatial resolution (submicronic) can be obtained, so that individual particles can be resolved, and a three-dimensional reconstruction of the particle's arrangement and the solvent crystals is possible, providing more information's than 2D observations from cross-sections.

The interest of these techniques is illustrated here through the in-situ observation of directional solidification of colloidal alumina suspensions. In this first part of the two companion papers, we focus on the initial instants of solidification. The other companion paper is focussed on the steady state behaviour.

**2. Experimental methods**

Slurries were prepared by mixing distilled water with a small amount (1 wt% of the powder) of ammonium polymethacrylate anionic dispersant (Darvan C, R. T. Vanderbilt Co., Norwalk, CT) and the alumina powder (TM-DAR from Taimei, Japan ($D_{50}$=0.2 microns, SSA 14.5m²/g). Slurries were ball-milled for 20 hrs with alumina balls and de-aired by stirring in a vacuum desiccator, until complete removal of air bubbles (typically 30 min).

Solidification experiments were performed by pouring the suspensions into of tube like shape (which allows a constant thickness to be crossed by the Xrays beam, i.e; with no significant contribution to X-rays data, in addition to a low absorption coefficient) placed at the surface of a cold finger, in a cell cooled using liquid nitrogen (Figure 1), available at the ESRF beamline. Freezing kinetics were controlled by a heater placed at the bottom of the cold finger (copper rod) and a thermocouple also placed at the bottom. The experimental setup allows only linear ramp of temperature during cooling. A cooling ramp of 5°C/min was used in the experiments, similar to the conditions of previous experiments [7,17,18].

The experimental part of this study focuses on the use of two and three-dimensional non-destructive techniques, i.e. high-resolution X-ray absorption radiography and tomography, to directly image nucleation and growth of the crystals in the suspension and the corresponding particles redistribution. Tomography combines information from a large number of X-ray radiographs taken with different viewing angles of the sample. The technique includes a computed step, i.e., a recalculation step during which a 3D map of the local absorption coefficients in each elementary volume of the sample is retrieved from the set of absorption radiographs. In this study, the HST program available at the ESRF was used [19]. The reconstructed map gives an indirect image of the 3D microstructure.

Although dramatic improvement in the acquisition time of tomography images have been made in the last decade, acquiring a full 180° scan cannot be obtained in less than approximately 30s, and such time span depends on the desired resolution; higher resolution acquisitions require 5 to 10 min. To obtain a reconstruction of good quality, the interface must remain at a constant location during acquisition. Hence, in situ tomography observations can only be obtained for very low interface velocities (<1 micron/s). Only one scan of in situ tomography of the moving interface during solidification was performed, the results of which are reported in the companion paper (part II: Steady State). Because of the typical solidification front velocities in these experiments, a different approach was adopted. Dynamics of solidification was followed by X-ray radiography, with an acquisition frequency of 3 Hz, which allows to precisely follow the interface evolution in two dimensions. The frozen structures after complete solidification were characterized in three dimensions afterwards using a low speed high-resolution tomography acquisition, and the results were then combined with the data obtained by radiography.

In the present study, the samples were scanned using a high-resolution X-ray tomograph located at the ESRF (beam line ID 19) in Grenoble (France). X-ray tomography was performed at a voxel size of $(1.4 \ \mu m)^3$. The energy was set to 20.5 keV. The distance between the sample and the detector was 20 mm. Because of the extremely high coherence of the X-ray beam on this beam line, absorption is not the only source of contrast in the obtained radiographs and phase contrast is also present, but in a small amount. A set of 1200 projections was taken within 180°. The detector was a CCD camera with $2048 \times 2048$ sensitive elements coupled with an X-ray-sensitive laser screen. To improve the acquisition frequency, the radiographs were acquired in a binning configuration (i.e. with a 1024x1024 pixels resolution, half that of tomography acquisitions). The corresponding spatial resolutions of the X-ray absorption radiography and tomography results presented here are given in table 1.

## 3. Results

### 3.1 Observation artefacts and defects

Several precautions must be taken to obtain high quality, artefacts-free images. First, any condensation from the atmosphere at the surface of the mold must be avoided. Such condensation would initiate growth of ice crystals at the surface of the mold, which would be detrimental for the observations. A special cooling cell was therefore developed (see again figure 1), including a double circulation of nitrogen at different temperatures, to avoid condensation at the surface of the mold, while ensuring that solidification remains directional. Second, the prolonged exposition of the suspension under the beam induces the formation of bubbles, which are rapidly growing with time.

Under typical experimental conditions, bubbles start to appear after 5 to 10 min of irradiation and rapidly grow. To avoid their presence during the image acquisition, every experiment was performed twice: once with a constant exposition under the beam, which allows the precise determination of the observation time frame (which depends on the experimental conditions). A second go was performed, switching on the X-ray beam (and hence the radiography acquisition) only a few seconds before the useful observation time frame. This experimental strategy is time consuming but ensures bubbles-free observations. The frozen structures are not affected by the beam during the tomography acquisitions, no particles movement is possible.

Finally, the effect of the size of the mold in which suspensions were poured was investigated. The signal is collected on a 2048 pixels wide detector. The mold inner diameter must fit into this dimension to get an optimal reconstruction of the data; therefore, the spatial resolution is dependent on the size of the sample to be imaged. The diameter of the mold must therefore be adjusted as a function of the desired spatial resolution. It has been verified in this study that similar results (in terms of solidified structure) were obtained with mold diameters in the range 0.6 – 6 mm. Side effects were never observed, even for the lowest diameter. For normal-resolution observations, a tube diameter of 3.2mm was used. For high-resolution observations (shown in the companion paper), a mold diameter of 0.6 mm was used.

**3.2 Initial instants of solidification**

The initial instants of solidification are shown in the pictures sequence in figure 2, obtained by radiography. The acquisition frequency is still too low to observe precisely the nucleation stage. In the time span (0.3s) between the last frame with no crystals and the next one, crystals have already grown around 250 microns in this example. The minimum growth velocity during these first instants can accordingly be estimated to be 750 microns/s, and is probably faster than that. The interface velocity can be measured from the pictures sequence. To limit the error range of the measurements, four sets of measurements were done, at various locations along the x-axis defined in figure 2. The results are plotted in figure 3. The crystals are first growing at a very high rate (>750 microns/s) that dramatically slows down within the first 4 seconds. An almost steady state is then observed, with a rate around 50 microns/s. Such profile of the interface velocity is not surprising considering the cooling conditions. A linear cooling ramp being applied, an inverse exponential variation of the velocity should be expected, as observed.

Secondary crystals, highly dendritic, can also be noticed in the initial stages, ahead of the main interface, in particular in figure 2b and 2c. These crystals seem to be growing only during the first nucleation and growth instants, then come to a halt and, before the main interface catches up, seem to melt back. Two explanations can be considered: these crystals can be either particularly well-oriented crystals (crystallographically speaking), growing initially very fast, or they can be crystals growing at the surface of the mold. In this latter case, the thermal conditions might be somewhat different from the ones at the copper surface, and very likely, nucleation might be easier. Their growth rapidly stops, the thermal conductivity of the polypropylene mold being low. The radiography pictures providing only a two-dimensional information, a definitive conclusion cannot be drawn from the current observations.

Using the tomography experiment performed on the frozen body, the solidified structure corresponding to the initial instants observed in figure 2 can be reconstructed. The corresponding cross-sections in the xz plane (parallel to the growth direction) and xy plane (perpendicular to the growth direction) are shown, respectively, in figure 4a and figures 4b-e. The z location of the xy cross-sections are indicated in figure 4a. The bottom part of figure 4a reveals the copper surface on which nucleation occurs. The ice crystals appear in dark in the pictures, while the lighter phase corresponds to the particle rich phase (concentrated entrapped particles between the ice crystals). A morphological transition is clearly visible, with an evolution of the crystals shape and arrangement from the copper surface to point c. In the initial stages (figure 4d and 4e), a dual microstructure is observed, with both vertical lamellar crystals (tagged 1) and almost horizontal lamellar crystals (tagged 2). The structure cannot be resolved within the first 80 microns, the spatial resolution achieved by tomography being too large compared to the typical size of the microstructure in this portion of the sample. A lamellar structure is observed in point c and from then on remains constant.

The attenuation contrast between the ice crystals and the concentrated particles being strong enough, a quantitative analysis of the phase fraction can be performed on the xy cross-sections, to investigate the evolution of this parameter along the growth direction. The results are plotted in figure 5. The plotted particle rich phase fraction corresponds to the fraction of the phase composed of concentrated particles and the ice surrounding these particles (hence, if all particles are entrapped with no redistribution taking place, the particle rich phase fraction defined here would be equal to 1), as opposed to the phase composed of pure ice crystals (particle-free phase)[†]. Five distinct areas can be identified on the graph. A first region A where the tomography resolution was not good enough to get reliable images. In region B , the particle rich phase fraction goes progressively down to a minimum. A second rapid increase of the particle rich phase fraction occurs in region C. The fraction is then progressively decreasing in region D, until becoming constant in region E, reaching a constant value around 0.49.

## 4. Discussion

---

[†] The actual particle fraction in the phase composed of the concentrated particles and the surrounding ice is theoretically close to 0.55-0.58 [20], depending on particle size.

The existence of a morphological transition of the interface in the initial stages of solidification under similar conditions has been proposed previously [7], to explain the evolution of the structure close to the cooling surface as revealed by SEM observations. However, the observations were only two-dimensional. From the qualitative and quantitative results presented here, the scenario can be greatly clarified.

It was previously suspected that the interface was initially flat and then undergoing a destabilization by a Mullins-Sekerka type mechanism [21], leading to a progressive evolution of the interface morphology from flat to cellular and finally lamellar. It is clear from the observations presented here that this is not the case. Instead, two types of crystals are present from the beginning: a first population of lamellar crystals with their main axis oriented along the z-axis (cooling direction, and hereafter called z-crystals), and a second population of crystals, more or less lamellar, called r-crystals (r for random, Figure 6), but lying more or less in the xy-plane (radial direction). The lenticular or lamellar shape of these crystals can be explained through the strong growth anisotropy of the ice crystals in the conditions of temperature and pressure used here. Growth along the *c* axis of hexagonal ice is theoretically two orders of magnitude slower than growth along the *a* and *b* axis, explaining their lamellar morphology. Hence, both types of crystals exhibit the same crystallography, but present a different orientation to the temperature gradient. The nucleation stage occurring at the surface of the copper rod should be spatially homogeneous (no epitaxy relationships between the copper surface and the ice); crystals with all crystallographic orientations are initially nucleating and growing, explaining the two populations of crystals observed. Although we cannot conclude completely regarding the presence of both types of crystals in the 0-80 microns zone, it seems reasonable to assume that they are both nucleating at the surface. It is also worth mentioning that the relative growth kinetics of the two types of crystals cannot yet be discriminated with radiography experiments. They can possibly be different, with one type of crystals growing faster than the other one.

The evolution of the ice crystals and particles redistribution is then dependent on several parameters, namely:

- The growth kinetics of the various types of ice crystals. z-crystals are growing much faster in the z direction than r-crystals.

- The ability and efficiency of these crystals to repel the particles and therefore induce particles redistribution in the suspension (figure 7). The particles redistribution efficiency of a given crystals interface section can be defined as the product of the interface surface of the considered portion by the interface velocity of this same portion. Particles are repelled in a direction tangential to the interface displacement direction. Only the tip of the z-crystals is redistributing particles in the z-direction in figure 7a, hence a low redistribution efficiency is found in the z-direction, while the opposite situation is encountered in figure 7b, with redistribution mostly occurring in the z direction. Any tilt of the crystals will result in an intermediate situation.

- The thermal conditions. In these experiments, solidification is directional; the suspension is continuously cooled from the bottom (starting from room temperature), so that a temperature gradient along the z-axis is progressively appearing.

- The suspension properties which depend on particles concentration. In particular, it has been previously shown that particles can only be concentrated up to a certain breakthrough point [20]. Particles will be pushed along the interface until the capillary drag force pushing the particles is countered by the force resulting from the osmotic pressure of the suspension [10,11,22,23]. The resulting critical particles packing fraction (0.55-0.58) was close to the maximum packing, which is itself dependent on the particles characteristics (size distribution, roughness, etc…). This aspect is particularly important for low interface velocities [24].

- The efficiency of particles packing between the growing crystals. When the particles dimensions and the crystals spacing are of the same order of magnitude, the optimal particles packing (defined in the previous point) will be more difficult to achieve (figure 8). We can define the true particles fraction in the particle rich phase by the following equation:

$$Cp_t = Cp_{max} \cdot P_{eff} \qquad \text{(equation 1)}$$

where $Cp_t$ is the true particles fraction, $Cp_{max}$ is the maximum particles fraction obtained at the breakthrough (when the osmotic force takes over the capillary force), and $P_{eff}$ is the packing efficiency. $P_{eff}$ is a coefficient that is always smaller than 1, a value of 1 corresponding to the situation where no packing faults are present, i.e. a situation of colloidal crystal. The lower limit of $P_{eff}$ is dictated by geometrical considerations (likely percolation arguments) and might be estimated at a value around 0.2 based on the percolation theory for spherical particles. When $P_{eff}=1$, then the value of $Cp_t$ corresponds to that obtained through the models mentioned earlier, and is around 0.55-0.58. The nominal particles fraction in the suspension is obtained by the product of the measured particle rich phase fraction (corresponding to the concentrated particles entrapped in ice) by $Cp_t$. We can verify here that with a measured value of $C_{pt}$ close to 0.48 at the end of the initial stage (Figure 5) and a nominal particle fraction of 0.28 (initial formulation of the suspension), we obtain a fraction of particles at breakthrough of 0.58, in excellent agreement with the model exposed previously.

Based on these parameters and the experimental qualitative and quantitative observations, the following scenario for the dynamical selection of interface pattern can be proposed:

- Zones A (defined as the portion of the sample where the tomography resolution was not good enough to image the crystals and particles distribution). When the temperature at the copper surface reaches the nucleation temperature, homogeneous (spatially speaking) nucleation of ice crystals occurs. Crystals with random crystallographic orientation are nucleating and growing very fast, so that all particles are initially entrapped. As the crystals start to grow along the z-axis, the interface velocity is diminishing, and particles redistribution can progressively take place, as particles start being repelled by the interface. All crystals with their c axis almost perpendicular to the temperature gradient align rapidly along the temperature gradient direction, so that their growth kinetics anisotropy corresponds to the favourable conditions dictated by the temperature gradient. This rapidly results in a situation where two populations of crystals are present: z-crystals and r-crystals. Since nucleation is spatially homogeneous at the surface of the cold finger, these populations are spatially homogeneously distributed, as seen in figure 9. The most favourable crystallographic orientation, as described previously, is also the one present in majority.

- Zone B. As the crystals grow larger and larger, particles' packing between the crystals becomes more and more efficient. The entrapped particles fraction therefore progressively diminishes. Crystals with all orientations grow along the z-axis. As the effect of the temperature gradient becomes more important, the growth kinetics along the z-axis slows down. r-crystals grow, repelling particles in the z–direction. The z-crystals do not affect the particles distribution in the z-direction, as they repell particles in the x and y directions (Figure 7). The coexistence of both types of crystals and their progressive growth are observed in figure 4d and 4e. The relative fraction (in the place) of z-crystals and r-crystals is plotted in Figure 5. The measured particle rich phase fraction corresponds to an average value of the two populations of crystals (Figure 10). The fraction of z-crystals progressively but slowly increases, corresponding to region B.

- Zone C. The r-crystals stop growing (or turn into z-crystals), as seen in figure 5, a much more favourable configuration under these directional solidification conditions. The influence of r-crystals on the particle rich phase fraction progressively diminishes to zero in region C; the particle rich phase fraction increases to its corresponding value for z-crystals only (figure 10). The extinction of r-crystals and preferential growth of z-crystals can be understood considering the thermal conductivity of the frozen part. Since the thermal conductivity of ice (1.6-2.4 W/mK, depending of temperature) and alumina particles (40 W/mK) largely differs (almost two orders of magnitude), we can estimate the apparent thermal conductivity in the z-crystals (assuming that thermal resistance of particle rich phase and ice act in parallel, the apparent thermal conductivity is close to that of alumina) and r-crystals domains (thermal resistance of particle rich phase and ice associated in series, the apparent thermal conductivity is assumed to be that of ice). Thus, the average thermal conductivity in z-crystals domain is larger than that in r-crystals domain (figure 11). After a certain time, the temperature ahead of the z-crystals domains is therefore lower than ahead of the r-crystals domain. Consequently, the r-crystals progressively stop growing. Particles concentrate above the stopping r-crystals, as shown in figure 7b and figure 12a-c, likely contributing to the short and sharp increase of the particle rich phase fraction.

- Zone D. Only z-crystals grow in the suspension, as seen in figure 4b-c; the concentrated particles above the recently stopped r-crystals still occupy a fraction of the xy-plane. It takes some time for the z-crystals or for the r-crystals turning into z-crystals to occupy that space (figure 12 d-g), so that these concentrated particles zones are progressively incorporated in the frozen structure. Particles redistribution is dictated by the orientation of the z-crystals to the growth direction. A progressive tilt of the crystals can be observed in figure 4a, corresponding to the alignment of the z-axis crystals to the direction of the temperature gradient. The particle rich phase fraction is dictated by the efficiency of particles packing, as discussed previously, but this parameter is now dependent only on one type of crystals (z-crystals) (Figure 10).

- Zone E. Particles redistribution occurs in the xy plane. The particles fraction $Cp_t$ is therefore constant, and is only dependent on the particle fraction initially present in the suspension. These aspects are discussed in more details in the companion paper (part II: Steady state). Spikes in the particle fraction and interface velocity in figure 5 can be noted, close to z=2100 microns. These spikes are significant (they do not fall within the measurement error range), and correspond indeed to interface instabilities, but this will be discussed in more details in a separate paper.

Several consequences arise from this analysis. First, the magnitude of particle redistribution locally depends on the orientation of the ice crystals to the growth direction. This has been proven with the different behaviour associated to the z- and r-crystals. Therefore, any modification of the crystals morphology should be locally reflected to some extent into the particle fraction distribution in the solidified body, and hence in the structure of the materials processed through this technique. This also implies that removing this initial transition zone, which could be detrimental for functional properties, will be difficult to achieve. It would imply favouring the z-crystals from the beginning.

This could be done by avoiding nucleation of the r-crystals, which could be achieved through a carefully selected templating of the cold surface with a substance exhibiting a crystal lattice correspondence (epitaxy) with the a-axis of the hexagonal ice structure. An alternative solution could be to suppress the strong anisotropy of growth kinetics with additives blocking certain growth directions, as encountered in nature [25]. Using a different material, or with different thermal properties, the dimensions of the initial transition zone should be affected accordingly.

The second important result here is the identification of several parameters affecting the efficiency of particle packing $P_{eff}$. The coefficient $P_{eff}$ is not only dependent on the size relationships between the particles and the available space for packing (i.e. the interlamellar spacing), but also on the orientation of the growing crystals. The global value of $P_{eff}$, which dictates the evolution of the measured particle fraction during the initial stages, can be separated here into two contributions: $P_{eff}$(r-crystals) and $P_{eff}$(z crystals). More generally, $P_{eff}$, is probably dependent on the angle $\theta$ between the local interface displacement direction (of the portion of the interface actually repelling the particles) and the temperature gradient. $P_{eff}$ can therefore be defined as:

$$P_{eff} = \sum_{\theta=0}^{90} f(\theta) P_{eff}(\theta) \qquad \text{(equation 2)}$$

where $f(\theta)$ is the fraction of crystals with the $\theta$ orientation, and $P_{eff}(\theta)$ is the corresponding packing efficiency. In addition, the results presented here (in particular Figure 5) reveal that $P_{eff}$(r-crystals) >$P_{eff}$(z crystals). This is justified by two facts: the orientation of the r-crystals might be more favourable for proper packing, and, more importantly, their size is larger than z-crystals, as seen in figure 6; less packing defects can therefore be expected, in agreement with the results shown in figure 5. When r-crystals stop growing, only the z-crystals are remaining, and the particle fraction is increasing to its new equilibrium value, dictated by $P_{eff}$(z-crystals). The scenario exposed here could explain the previously mentioned influence of the particles dimension over the dimensions of this initial transition zone [18]. Larger particles lead to transition zone spreading over longer distances. Since the size of the transition is largely dependent on the particle redistribution behaviour, the redistribution is expected to be more difficult with larger particles, as it takes more time for $P_{eff}$ to reach its maximum value (i.e. for a given interlamellar spacing, particles packing is more efficient with smaller particles).

The third consequence of the presence of lamellar crystals from the beginning is the possibility to use templated surfaces to induce ordered nucleation at the surface, leading to lamellar structures with long range order in the xy plane.

**5. Conclusions**

Based on the experimental in situ investigations by X-ray radiography and tomography of the controlled solidification of concentrated alumina particles suspensions, the following conclusions can be drawn, regarding the initial stages of solidification and the dynamical selection of interface pattern:

- Nucleation of ice crystals is occurring homogeneously (spatially speaking) at the cold surface, leading to two types of lamellar crystals, z-crystals and r-crystals, both corresponding to crystals growing perpendicular to the c-axis of the hexagonal ice structure, but with a different orientation in regard to the temperature gradient.

- Particle redistribution in general and in particular during these first stages is different for the different types of crystals growing into the suspensions. z-crystals growing vertically are redistributing particles in the xy plane (perpendicular to the solidification direction), while r-crystals are repelling particles in xz and yz planes (parallel to the solidification direction). The relative importance and evolution of both populations of crystals is reflected in variations of the fraction of particles entrapped in the solidified structure. Particles' packing between the growing crystals is becoming progressively more efficient up to a distance where the r-crystals stop growing, leaving only z-crystals growing into the suspension.

- The efficiency of particles packing, defined as the ratio of the maximum particles concentration achievable and the actual particle fraction obtained, is dependent on the ratio of particle size and intercrystals (or lamellar) spacing, and the orientation relationships between the ice crystals and the temperature gradient.

**Acknowledgements**

We acknowledge the European Synchrotron Radiation Facility for provision of synchrotron radiation beam time and we would like to thank Elodie Boller and Jean-Paul Valade for their irreplaceable assistance in using beamline ID19. Financial support was provided by the National Research Agency (ANR), project NACRE in the non-thematic BLANC programme.

**References**

[1] H. Ishiguro and B. Rubinsky, "Mechanical interactions between ice crystals and red blood cells during directional solidification," *Cryobiology,* **31**[5] 483-500 (1994).
[2] H. F. Zhang, I. Hussain, M. Brust, M. F. Butler, S. P. Rannard, and A. I. Cooper, "Aligned two- and three-dimensional structures by directional freezing of polymers and nanoparticles," *Nature Materials,* **4**[10] 787-793 (2005).
[3] P. Viklander, "Laboratory study of stone heave in till exposed to freezing and thawing," *Cold Regions Science and Technology,* **27**[2] 141-152 (1998).
[4] G. Gay and M. A. Azouni, "Forced migration of nonsoluble and soluble metallic pollutants ahead of a liquid-solid interface during unidirectional freezing of dilute clayey suspensions," *Cryst. Growth Des.,* **2**[2] 135-140 (2002).

[5] S. Deville, "Freeze-Casting of Porous Ceramics: A Review of Current Achievements and Issues," *Adv. Eng. Mater.,* **10**[3] 155-169 (2008).

[6] Y. Chino and D. C. Dunand, "Directionally freeze-cast titanium foam with aligned, elongated pores," *Acta Mater.,* **56**[1] 105-113 (2008).

[7] S. Deville, E. Saiz, and A. P. Tomsia, "Ice-templated porous alumina structures," *Acta Mater.,* **55** 1965-1974 (2007).

[8] G. Lipp and C. Korber, "On the engulfment of spherical particles by a moving solid liquid interface," *J. Cryst. Growth,* **130** 475-489 (1993).

[9] J. Cisse and G. F. Bolling, "A study of the trapping and rejection of insoluble particles during the freezing of water," *J. Cryst. Growth,* **10** 67-76 (1971).

[10] C. Korber, Rau, G., Cosman, M.D., Cravalho, E.G., "Interaction of particles and a moving ice-liquid interface," *J. Cryst. Growth,* **72** 649-662 (1985).

[11] R. Asthana and S. N. Tewari, "Engulfment of foreign particles by a freezing interface," *J. Mater. Sci.,* **28**[20] 5414-5425 (1993).

[12] S. S. L. Peppin, J. A. W. Elliot, and M. G. Worster, "Solidification of colloidal suspensions," *J. Fluid Mech.,* **554** 147-166 (2006).

[13] S. S. L. Peppin, M. G. Worster, and J. S. Wettlaufer, "Morphological instability in freezing colloidal suspensions," *Proc. R. Soc. London, Ser. A,* **463**[2079] 723-733 (2007).

[14] E. Maire, J. Y. Buffière, L. Salvo, J. J. Blandin, W. Ludwig, and J. M. Letang, "On the application of X-ray microtomography in the field of materials science," *Adv. Eng. Mater.,* **3**[8] 539-546 (2001).

[15] L. Salvo, P. Cloetens, E. Maire, S. Zabler, J. J. Blandin, J. Y. Buffière, W. Ludwig, E. Boller, D. Bellet, and C. Josserond, "X-ray micro-tomography an attractive characterisation technique in materials science," *Nuclear Instruments and Methods in Physics Research, Section B: Beam Interactions with Materials and Atoms,* **200**[SUPPL.] 273-286 (2003).

[16] J. A. Sekhar and R. Trivedi, "Solidification microstructure evolution in the presence of inert particles," *Mater. Sci. Eng., A,* **A147**[1] 9-21 (1991).

[17] S. Deville, E. Saiz, R. K. Nalla, and A. P. Tomsia, "Freezing as a path to build complex composites," *Science,* **311** 515-518 (2006).

[18] S. Deville, E. Saiz, and A. P. Tomsia, "Freeze casting of hydroxyapatite scaffolds for bone tissue regeneration," *Biomaterials,* **27** 5480-5489 (2006).

[19] HST Program, http://www.esrf.eu/computing/scientific/FIT2D/HST/hst.html

[20] N. O. Shanti, K. Araki, and J. W. Halloran, "Particle Redistribution During Dendritic Solidification of Particle Suspensions," *J. Am. Ceram. Soc.,* **89**[8] 2444-2447 (2006).

[21] W. W. Mullins and R. F. Sekerka, "Stability of a planar interface during solidification of dilute binary alloy," *J. Appl. Phys.,* **35**[2] 444-451 (1964).

[22] A. W. Rempel and M. G. Worster, "The interaction between a particle and an advancing solidification front," *J. Cryst. Growth,* **205**[3] 427-440 (1999).

[23] M. A. Azouni and P. Casses, "Thermophysical properties effects on segregation during solidification," *Adv. Colloid Interface Sci.,* **75**[2] 83-106 (1998).

[24] S. S. L. Peppin, J. S. Wettlaufer, and M. G. Worster, "Experimental verification of morphological instability in freezing aqueous colloidal suspensions," *Phys. Rev. Lett.,* **100**[23] (2008).

[25] J. A. Raymond, P. Wilson, and A. L. DeVries, "Inhibition of Growth of Nonbasal Planes in Ice by Fish Antifreezes," *Proc. Natl. Acad. Sci.,* **86**[3] 881-885 (1989).

**Figures**

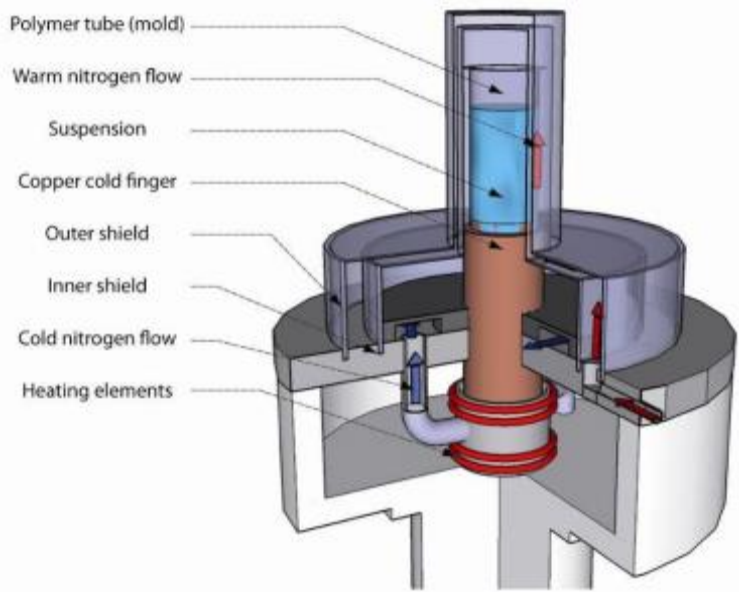

Figure 1: Cooling cell available at the ID19 beamline at the ESRF for the in situ experiments, to perform directional solidification while avoiding condensation at the surface of the cell.

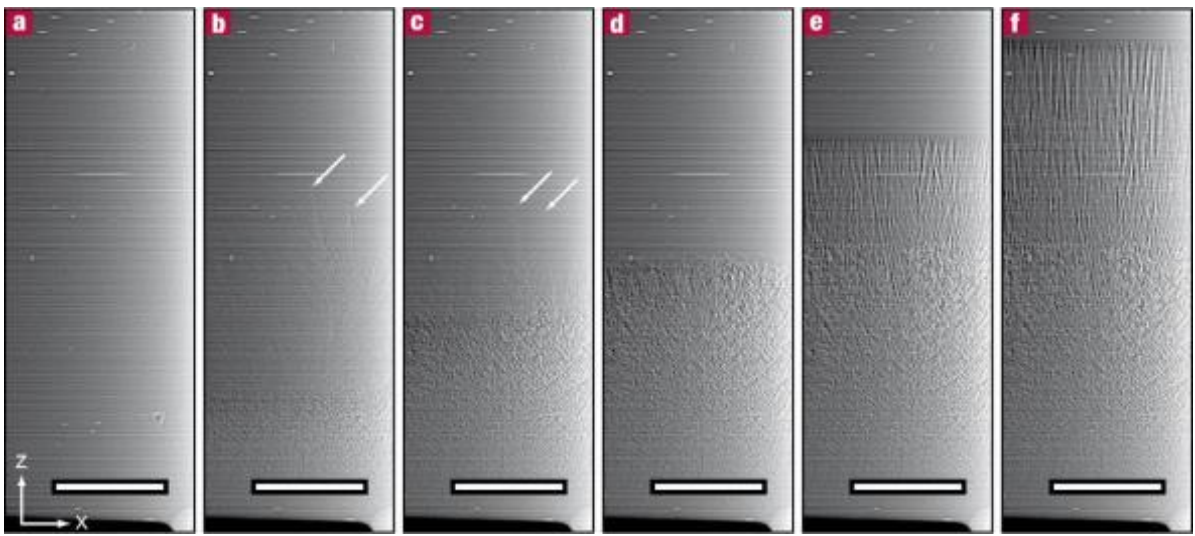

Figure 2: Solidification sequence in the initial instants, particle size 0.3 microns. Time picture was taken: a=0s, b=1s, c=2.1s, d=3.6s, e=9.7s, f=16.3s. Scale bar: 500 microns. Picture *a* is the last frame taken before nucleation and growth begin. The copper surface is visible in black at the bottom of the pictures. The white arrows indicate fast growing crystals.

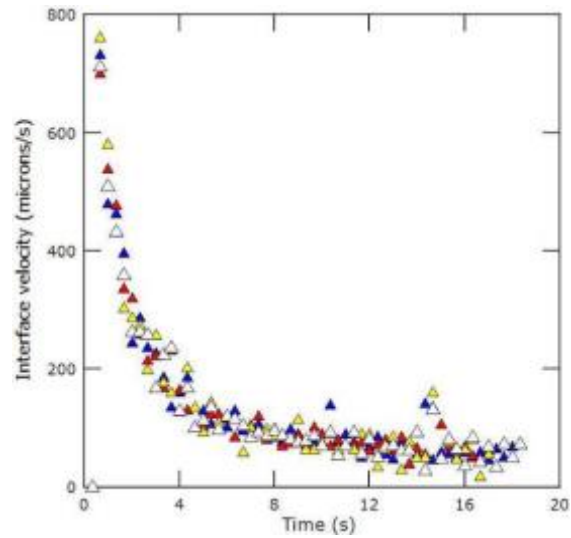

Figure 3: interface velocity vs. time in the first frozen zone, measured from the radiography images shown in figure 2, particle size 0.3 microns. The velocities values comprised four sets of measurements, performed at different location of the sample along the x-axis of the radiographs (see Fig. 2 for axis definition).

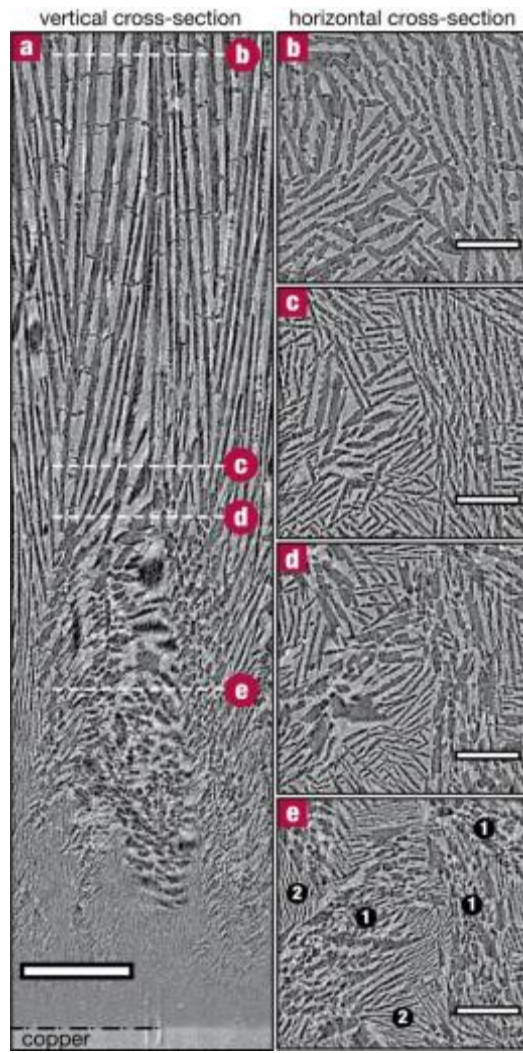

Figure 4: Tomography reconstructed slice showing the structure in the first frozen zone, vertical (left, perpendicular to solidification front) and horizontal (right, parallel to solidification front) cross-sections, particle size 0.3 microns. Scale bars: a : 250 microns, b-e : 150 microns. The location of the b-e cross-sections is shown in the micrograph *a*. The dashed line in *a* indicates the location of the copper surface, where nucleation occurs. Legend of panel e: zone 1: r-crystals, zone 2: z-crystals.

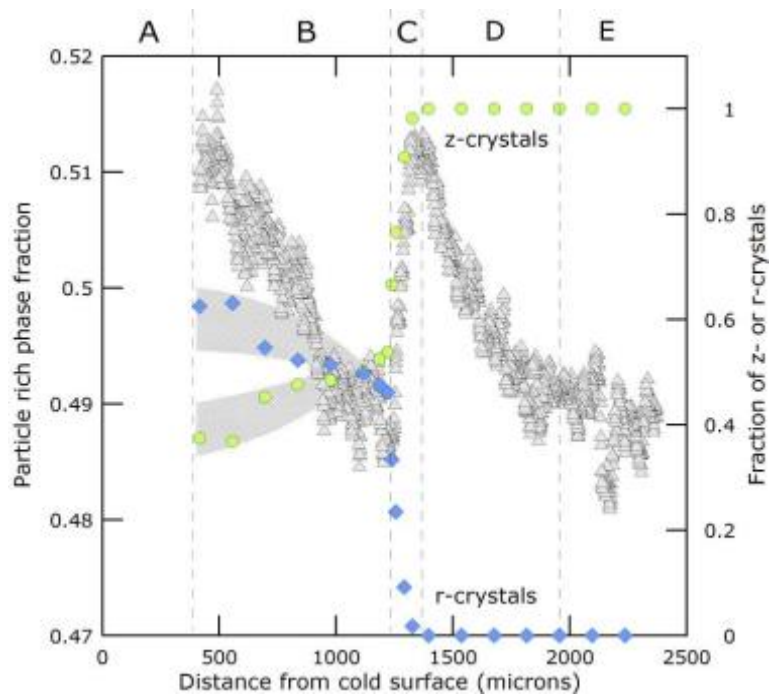

Figure 5: Particle rich phase fraction (▲) and fraction of z-crystals (●) and r-crystals (♦) (F(z-crystals)+F(r-crystals)=1, since only these two types of crystals are present) vs. distance from cold surface in the first frozen zone, measured from radiography and tomography data, particle size 0.3 microns. Dashed areas in region B indicate the measurement range of error, which increases as the distance from the cold surface decreases. The resolution was not good enough to obtain data in the region 0-500 microns (zone A). A similar particle fraction variation profile was obtained for all experiments, with particles fraction at the beginning of zone B up to 0.7-0.8, depending on particle size.

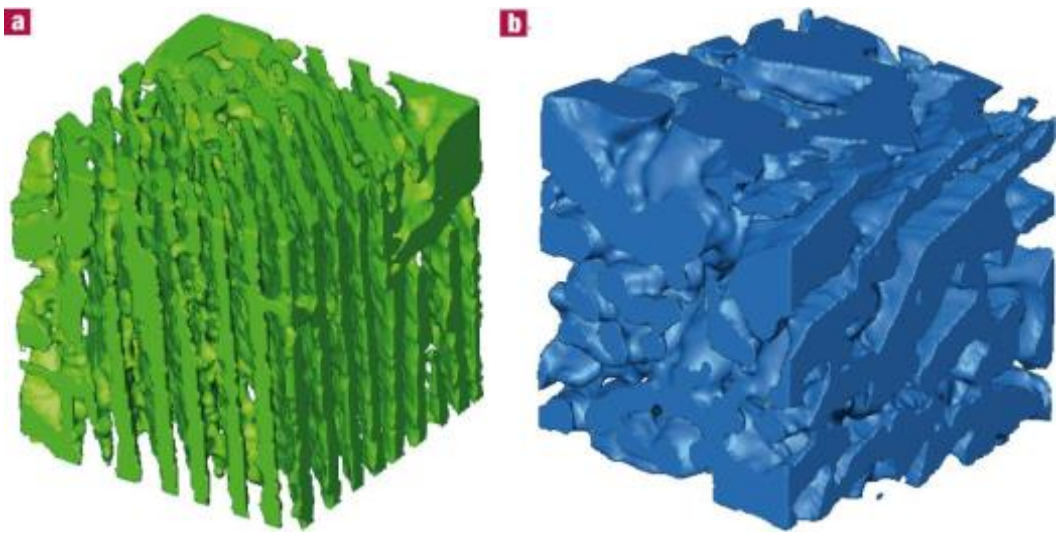

Figure 6: z-crystals (left) and r-crystals (right). Three-dimensional reconstruction from the tomography data, bounding box dimensions: 160x160x160 microns$^3$. The r-crystals have a larger interspacing.

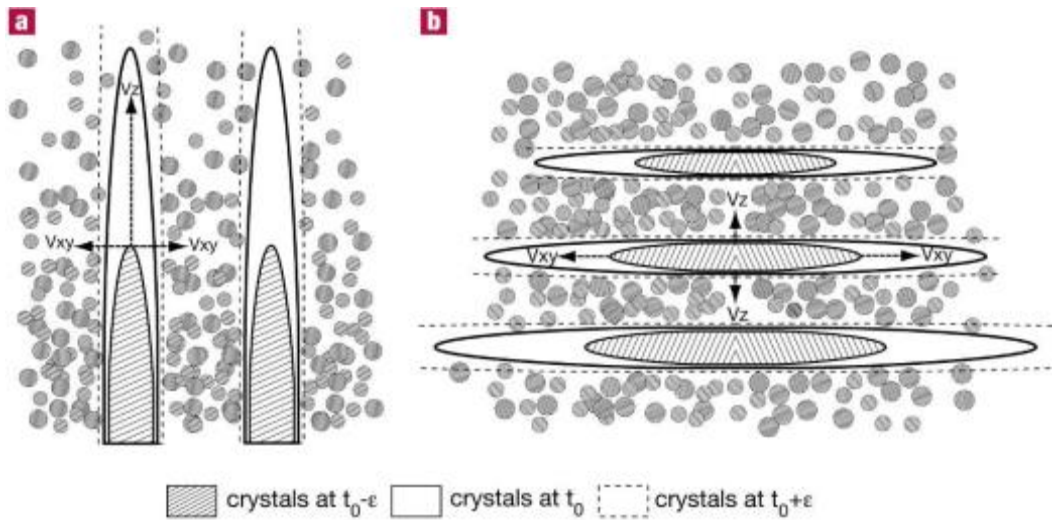

Figure 7: Schematic of particles redistribution as a function of the crystals orientation. (a) z-crystals ($\theta=90°$), where particles redistribution occurs in the xy plane and (b) r-crystals ($\theta=0°$), where particles redistribution occurs in the xz and yz planes. Dashed area: crystals at time $t_0-\varepsilon$, continuous line: crystals at time $t_0$, dashed line: crystals at time $t_0+\varepsilon$. Particles are repelled in a direction tangential to the interface displacement direction. Only the tip of the z-crystals in redistributing particles in the z-direction in figure a, while the opposite situation is encountered in figure b.

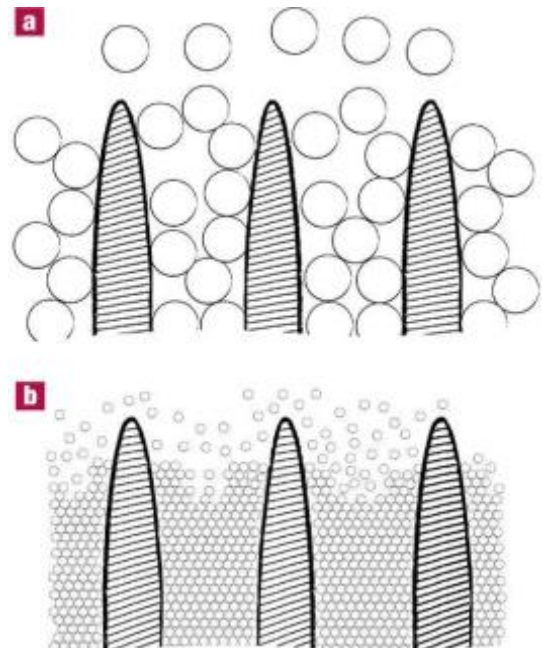

Figure 8: Schematic of the efficiency of particles packing for different particle size. With large particles (a), obtaining the maximum packing fraction is more difficult to achieve than with small particles (b).

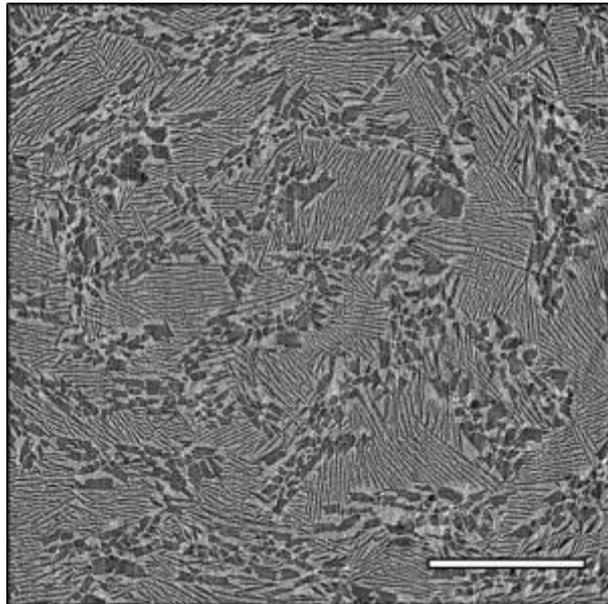

Figure 9: Cross-section in the xy-plane from tomography data showing the distribution of crystals in the initial stage of solidification. The majority of the crystals are z-crystals (lamellar shape in this cross-section) with their c axis in the xy plane. Scale bar 500 microns. These cross-sections can be used to measure the relative fraction of z- and r-crystals (see figure 5).

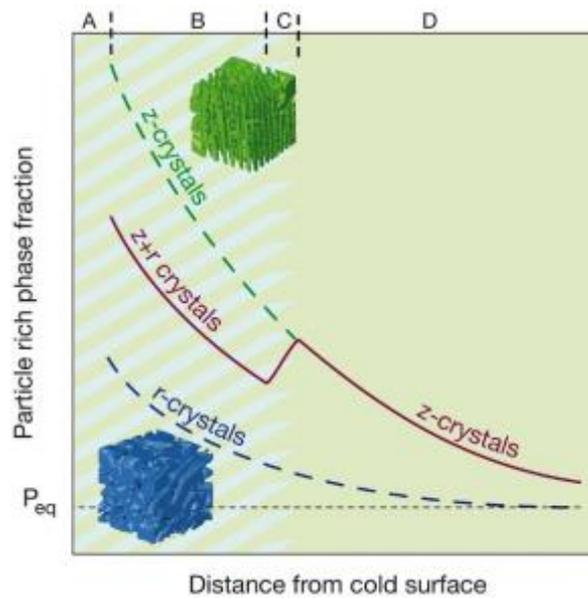

Figure 10: Schematic plot of particle rich phase fraction vs. distance from the cold surface, expected for the r-crystals (lower curve), z-crystals (upper curve) and their mixture (intermediate curve). r-crystals exhibit a better packing efficiency of particles in the particle rich phase, hence a lower particle rich phase fraction. When both types of crystals coexist (region B), an average particle rich phase fraction is obtained. When r-crystals stop growing (region C), the particle rich phase fraction is increasing to its corresponding value for z-crystals only. An equilibrium situation, where the particle rich phase fraction is equal to $P_{eq}$, is reached in any case; the value of $P_{eq}$ is given by the particles concentration at breakthrough and the nominal particle fraction in the suspension. It takes a longer time for the particle rich phase fraction to reach its equilibrium value in the case of z-crystals.

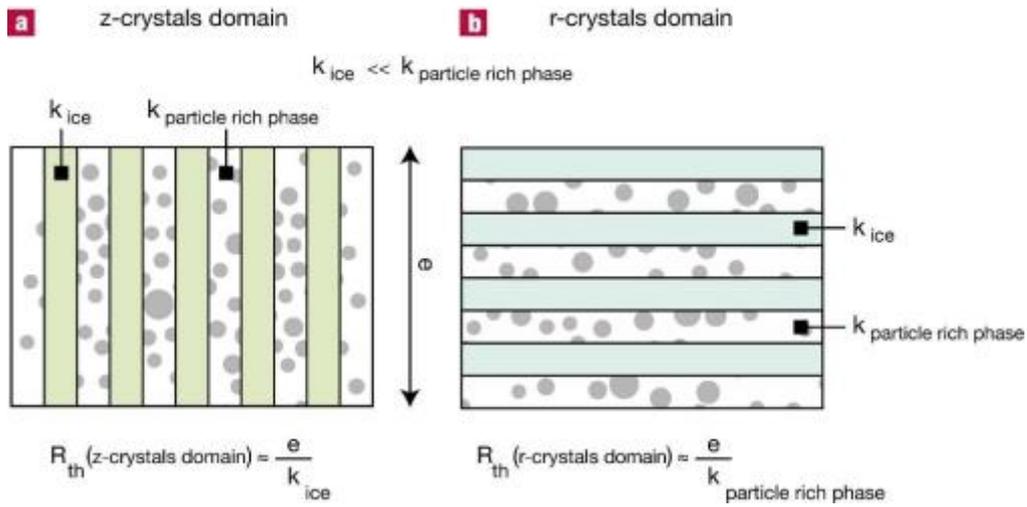

Figure 11: Thermal conduction and resistance for the r-crystals domain and z-crystals domain. The thermal conductivity of the z-crystals domain is larger; the temperature ahead of these domains should therefore be lower.

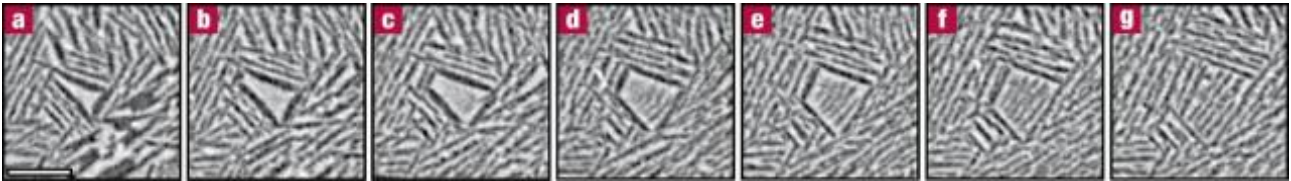

Figure 12: Cross-sections in the xy-plane from tomography data showing the end of transition zone. Detail showing the crystals evolution, with z-crystals progressively replacing r-crystals. Scale bar 100 microns. Z position: a:$z_0$, b: $z_0+39\mu m$, c: $z_0+60\mu m$, d: $z_0+80\mu m$, e: $z_0+111\mu m$, f: $z_0+129\mu m$, g: $z_0+151\mu m$. Pictures a to c are in zone C, d to g in zone D (reference zone defined in figure 5).

|  | Radiography | Tomography |
| --- | --- | --- |
| Normal resolution | 2.8 microns/pixel | 1.4 microns/voxel |
| High resolution | 0.56 microns/pixel | 0.28 microns/voxel |

Table 1: Spatial resolution (voxel or pixel size) obtained for radiography and tomography data. High resolution observations are shown in the companion paper (part II).